\documentclass[aps,prl,groupedaddress,fleqn,twocolumn]{revtex4-1}
\pdfoutput=1

\usepackage[pdftex]{graphicx}
\usepackage{amsmath}
\usepackage{amssymb}
\usepackage{gensymb}

\newcommand{\hl}[1]{#1}

\begin{document}

\title{Synchronization of two DFB lasers using frequency-shifted feedback for microwave photonics}

\author{Aur\'elien Thorette$^*$}
\author{Marco Romanelli}
\author{Marc Vallet}

\address{{\small Univ. Rennes, CNRS, Institut FOTON - UMR 6082, F-35000 Rennes, France} \\$^*$Corresponding author: aurelien.thorette@univ-rennes1.fr}

\maketitle

\textsc{Abstract:} The phase synchronization of two semiconductor DFB lasers submitted to both frequency-shifted cross-injection and self-feedback is investigated, both numerically and experimentally. A dedicated setup based on monolithic DFB lasers permits to effectively lock the frequency difference of the two lasers on an external reference over a broad range of parameters. In agreement with a delayed rate equation model, it is shown that the sensitivity of the locking to the optical feedback phases is mitigated by using strong cross-injection, thus paving the way to applications for microwave photonics.  Other synchronization dynamics can also be observed, such as bounded-phase oscillations.

\section{Introduction}


Phase synchronization of lasers is at the same time a widely used technique, and an active research domain, with applications in novel telecommunications and metrology. It is also a playground for the study of dynamical systems and generic synchronization. Laser synchronization experiments are often based on two well-understood mechanisms, namely optical injection and delayed self-feedback~\cite{wieczorek2005, erneux2010, sciamanna2015}. Combination of injection and feedback can lead to various phenomena~\cite{nizette2004}. For instance, it has been proposed as a way to sharpen the optical linewidth~\cite{simpson2013}. It has also been used to generate sidebands in the microwave domain, using so-called period-one dynamics~\cite{hung2015,chang2017}. Also, lasers can be forced into chaotic behavior while remaining synchronised~\cite{roy1994}, a property which opens interesting perspectives for chaotic communications~\cite{rogister2001}. More complex systems, involving a large number of coupled lasers have been investigated, with interests ranging from artificial neural networks to non-linear computing and signal processing~\cite{soriano2013}.


It has already been shown that frequency-shifted feedback (FSF) can be used to control and stabilize the frequency difference between the two polarization modes of  dual-frequency diode-pumped solid-state lasers, by injecting one mode into the other~\cite{kervevan2007, thevenin2011a}. This robust way to lock the beat-note phase on an external reference allows to generate stable optically carried microwave signals. It has also been used as a way to amplify the feedback signal for imaging applications~\cite{hugon2011}. FSF can also lead to more complex dynamics, such as bounded-phase phenomena~\cite{thevenin2011, romanelli2014}, excitability~\cite{romanelli2016} and the peculiar regime of bounded-phase chaos~\cite{thorette2016}. In this paper, we are interested in transposing the above technique to two separate semiconductor lasers, with the perspective of usages in microwave photonics systems, such as telecommunications and lidar-radar applications. 
\hl{However, given the lifetime of carriers and optical cavity field in semiconductor lasers (typically ranging from tens to hundreds of ps), in general the feedback delay time cannot be neglected, even in compact setups involving photonic integrated circuits}~\cite{primiani2016}. \hl{This applies even more to the setup studied in the present paper, where a 16 m long fiber feedback loop is present.}
Thus, this delay is expected to play a key role in the synchronization ability of the system~\cite{soriano2013}.
\hl{We also note that it is not always desirable to reduce the delay as much as possible. Indeed, a short delay may be attractive for integration purposes, however it is known that a km-long delay improves the stability and performances of opto-electronic oscillators}~\cite{vallet2016}. 

The behavior of two lasers subjected to both delayed feedback and injection has been studied before, analytically~\cite{flunkert2011}, numerically~\cite{hicke2011} and experimentally~\cite{liao2013}, showing phase locking, limit cycle dynamics, and chaotic waveforms. 
At variance with previous works, here we propose a numerical and experimental study of the locking mechanisms under the influence of frequency-shifted feedback, including a frequency difference between the lasers. We show this can lead to the synchronization of the beat-note frequency between the lasers on an external microwave reference.

\hl{This work is done with a microwave photonics perspective, as beatnote stabilization is a key requirement for the realization of heterodyne oscillators. Indeed, such systems provide an interesting alternative to direct modulation of light}~\cite{yao2009}. \hl{They have various advantages. First, they directly generate a single-sideband signal over an optical carrier, that is inherently insensitive to dispersion in a fiber link. They also display a 100\% modulation depth. Most importantly, they are broadly tunable and allow to reach high microwave frequencies (100~GHz). Finally, their microwave output can be locked on electronic local oscillators at lower frequency (e.g. 10 GHz) using various downconversion techniques}~\cite{rolland2011, pillet2014}.  

The paper is organized as follows. Section 2 presents the experimental setup, and the corresponding rate-equation model describing two lasers submitted to both cross-injection and self-feedback. Section 3 is devoted to the analysis and comparison of the numerical and experimental results. Finally, section 4 is dedicated to the conclusion and offers some perspectives.

\section{System and modeling}

\subsection{Experimental setup}

\begin{figure}[h!]
	\centering
	\includegraphics{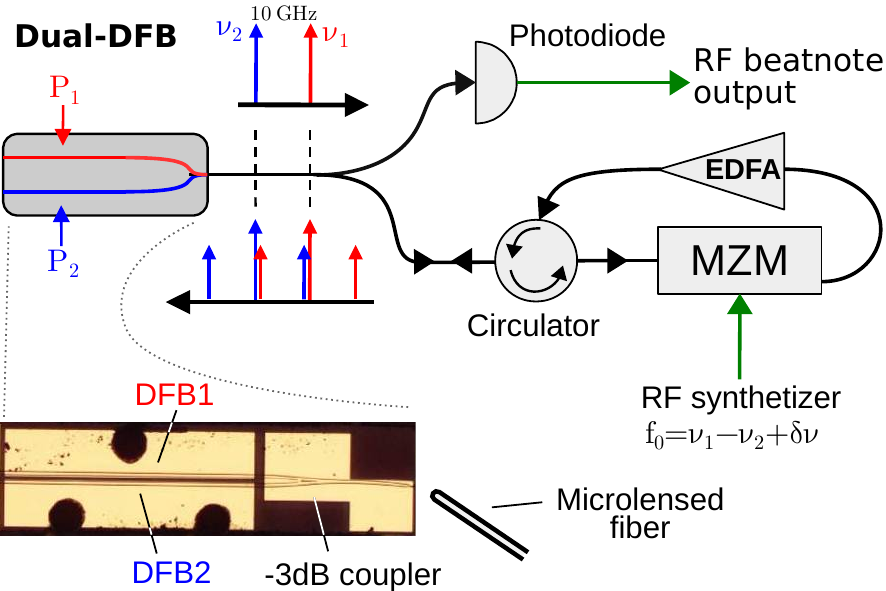}
	\caption{Experimental setup. See text for details.}
	\label{setup}	
\end{figure}

{The experimental setup, based on two DFB lasers, is shown in Fig.}~\ref{setup}. In order to minimize the uncorrelated environmental drifts, the two DFB lasers are part of the same custom InGaAsP photonic integrated component, which is described in detail in~\cite{vandijk2011}. \hl{One of its main features is a sharp optical linewidth of $300~\mathrm{kHz}$ for the two lasers, leading to a coherence length of about 1 km}. The two DFB lasers operate around $1.5~\mathrm{\mu m}$, but their optical frequencies can be continuously tuned by adjusting their pump current. This allows to select a frequency difference $\nu_1-\nu_2$ from 0 to few tens of GHz. Their outputs are combined by an on-chip coupler, and the $1~\mathrm{mW}$ of emitted light is collected by a microlensed fiber. Then, it travels through a \hl{16 m long} all-fibered feedback loop that contains a Mach-Zender modulator (MZM) and an amplifier (EDFA). A part of the signal is also routed to a photodiode, which records the beatnote between the different optical frequencies. This output signal in the microwave domain is then monitored on an electrical spectrum analyzer (ESA) and on an oscilloscope.

\par As seen on Fig.~\ref{setup}, the \hl{amplitude} modulator is driven by a reference oscillator at the microwave frequency $f_0$, and creates sidebands around each laser's optical frequency. When $f_0$ is close to the free-running frequency difference $\nu_1-\nu_2$, one of the sidebands becomes resonant for the other laser, which leads to cross-injection between the lasers. The cross-injection strength is controlled by the modulation rate $m$, that we choose to define as { $\mathcal{E}_\mathrm{out} = t_0 \, \mathcal{E}_\mathrm{in}\left[\sqrt{1-m} + \sqrt{\frac{m}{2}}\left(e^{2i\pi f_0 t} + e^{-2i\pi f_0 t}\right)+...\right] $. $\mathcal{E}_\mathrm{in}$ and $\mathcal{E}_\mathrm{out}$ are the input and output fields of the modulator respectively, and the coefficient $t_0<1 $ takes into account the total transmission and optical power associated to the higher harmonics. \hl{In the experiment, $m$ is controlled by changing the bias point of the MZM}.
\subsection{Rate equations model}

\begin{figure}[h!]
	\centering
	\includegraphics{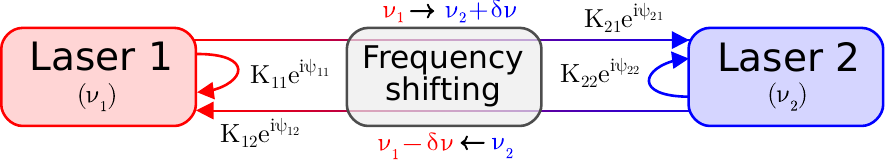}
	\caption{Coupling mechanisms between the two lasers.}
	\label{schema}	
\end{figure}

In order to model the experimental setup of Fig.~\ref{setup}, we consider two lasers coupled by self-feedback and cross-injection, as schematically sketched in Fig.~\ref{schema}. {On the injection path between the lasers, a frequency-shifting element is inserted, which brings each injecting field frequency closer to the free-running frequency of the injected laser, by shifting it by $\pm f_0$. The remaining frequency detuning is denoted as $\pm\delta\nu$}.
We use standard rate equations for class-B lasers~\cite{erneux2010} and include for each laser delayed self-feedback and cross-feedback terms. The equations for the intracavity fields $\mathcal{E}_j$ and normalized population inversions $N_j$ are:


\begin{subequations}
	\label{model}
	\begin{align}
	\begin{split}
	\frac{\mathrm{d}\mathcal{E}_1}{\mathrm{d}t}  =& \left(-\frac{1}{2\tau_p} +(1+i\alpha) g N_1 \right) \mathcal{E}_1 + 2i\pi\nu_1\mathcal{E}_1  \\ & \,\,+ K_{11} e^{i\psi_{11}} \mathcal{E}_1(t-T) + K_{12} e^{i\psi_{12}} \mathcal{E}_2(t-T)e^{2i\pi f_0 t}
	\end{split}\\
	\begin{split}
	\frac{\mathrm{d}\mathcal{E}_2}{\mathrm{d}t}  =& \left(-\frac{1}{2\tau_p}+ (1+i\alpha) g N_2 \right)\mathcal{E}_2 + 2i\pi\nu_2\mathcal{E}_2  \\ & \,\,+ K_{22} e^{i\psi_{22}} \mathcal{E}_2(t-T) + K_{21} e^{i\psi_{21}} \mathcal{E}_1(t-T)e^{-2i\pi f_0 t}
	\end{split}\\
	\frac{\mathrm{d}N_1}{\mathrm{d}t}  =& -\frac{N_1}{\tau_c} - 2gN_1|\mathcal{E}_1|^2 + P_1 \\
	\frac{\mathrm{d}N_2}{\mathrm{d}t}  =& -\frac{N_2}{\tau_c} - 2gN_2|\mathcal{E}_2|^2 + P_2
	\end{align} 
\end{subequations}

Here $\alpha$ is the linewidth enhancement factor, $g$ is the gain coefficient, $\tau_p$ and $\tau_c$ are the photon and carrier lifetime respectively. As we consider two identical lasers, these values are supposed to be the same for both lasers, leading to equal threshold pumping current $P_\mathrm{th}$.  However, we allow the lasers to have different frequencies $\nu_j$ and pumping currents $P_j$. The $K_{ij}$ coefficients quantify the injection strengths, and the associated angle $\psi_{ij}$ are additional phase shifts, introduced phenomenologically to account for erratic phase drifts of thermal and/or mechanical origin in the feedback loop, and dispersion effects in fiber, that may play a role in a real experiment. They also include any constant phase shift introduced by the modulator. 
\hl{As we are interested in determining the range of parameters allowing for phase-locking, and not in reproducing the phase noise spectrum of the system, noise terms have not been considered in the model. The absence of noise implies that the coherence length is infinite in the model, leading to coherent feedback and injection irrespective of the delay. Given the ratio between the feedback loop length (16 m) and the coherence length (1 km), this assumption matches well with our experimental conditions.}
We introduce equations for the normalized slowly varying amplitudes of the fields by using the rotating frames $\mathcal{E}_1 = E_1 e^{-i\xi} e^{2i\pi\nu_1 t}$ and $\mathcal{E}_2 = E_2 e^{2i\pi(\nu_1-f_0) t}$ with arbitrary constant phase $\xi$, along with normalizations $e_j=E_j/\widehat{E_j}$ and $m_j=(N_j-N_j^0)/\widehat{N_j}$. We also introduce the pumping rates $r_j=P_j/P_\mathrm{th}$ and normalized damping coefficient $\varepsilon' = \sqrt{\frac{\tau_p}{\tau_c(r_1-1)}}$. In order to obtain non-stiff equations, we choose a time scale $s=2\pi f_R^{(1)} t$, which is normalized to the frequency of the relaxation oscillations of the first laser $f_R^{(1)}=\frac{1}{2\pi}\sqrt{\frac{r_1-1}{\tau_c\tau_p}}$. This leads to normalized delay $\tau = 2 \pi f_R^{(1)} T$, detuning $\delta = (\nu_1-\nu_2-f_0)/f_R^{(1)}$, and injection strengths $\kappa_{ij}= K_{ij}/ 2\pi f_R^{(1)}$.
The equations for the slowly varying amplitudes $e_j$ and populations $m_j$ read:

\begin{subequations}
	\label{rmodel}
	\begin{align}
	\begin{split}
	\frac{\mathrm{d}e_1}{\mathrm{d}s}  =& (1+i\alpha) \frac{m_1 e_1}{2} + \kappa_{11} e^{i\varphi_{11}} e_1(s-\tau) \\&\hspace{2.1cm} + \kappa_{12} e^{i\varphi_{12}}  e_2(s-\tau) 
	\end{split} \\	
	\begin{split}
	\frac{\mathrm{d}e_2}{\mathrm{d}s}  =& (1+i\alpha) \frac{m_2 e_2}{2} - i\delta e_2 + \kappa_{22} e^{i\varphi_{22}} e_2(s-\tau) \\&\hspace{3.15cm} +\kappa_{21} e^{i\varphi_{21}} e_1(s-\tau) 
	\end{split} \\
	\label{pop1}
	\frac{\mathrm{d}m_1}{\mathrm{d}s}  =& 1 - |e_1|^2 - \varepsilon' m_1 \left(1+(r_1-1)|e_1|^2\right) \\
	\label{pop2}	
	\frac{\mathrm{d}m_2}{\mathrm{d}s}  =& \frac{r_2-1}{r_1-1} - |e_2|^2 - \varepsilon' m_2 \left(1+(r_1-1)|e_2|^2\right)
	\end{align} 
\end{subequations}


In the amplitude equations, we see the appearance of the following feedback phases $\varphi_{ij}$, arising from the propagation of the optical waves in the feedback loop of length $L = L'+L''$ (with $L'$ the length from the lasers to MZM, and $L''$ the distance from the MZM to the lasers), assuming optical frequencies $\nu_1$ and $\nu_1-f_0$, respectively.

\begin{subequations}
	\label{phases}
	\begin{align}
		\varphi_{11}  &= -2\pi \frac{\nu_1 nL}{c} +\psi_{11} \equiv \varphi_1  \\
		\varphi_{12}  &= -2\pi\frac{(\nu_1-f_0) nL' + \nu_1 nL''}{c} +\psi_{21}-\xi  \equiv\varphi_x \\	
		\varphi_{21}  &= -2\pi\frac{\nu_1 nL' + (\nu_1-f_0) nL''}{c} +\psi_{12}+\xi
		\equiv\varphi_x-\delta\tau \\
		\varphi_{22}  &= -2\pi \frac{(\nu_1-f_0) nL}{c}+\psi_{22} \equiv \varphi_2-\delta\tau \\
		2\xi &= 2\pi\frac{f_0 n (L'-L'')}{c}+\psi_{21}-\psi_{12}-\delta\tau
	\end{align} 
\end{subequations}

Using an appropriate choice of $\xi$ (see Eq.~(\ref{phases}e)), we can identify only three phases of physical significance $\varphi_1$, $\varphi_2$ and $\varphi_x$ that do not depend on the detuning. These phases cannot be controlled experimentally, and are prone to fluctuate due to the frequency drift of the lasers, or to variations of the fiber optical path induced by environmental fluctuations.   

\par{Compared to the scheme of Fig.~\ref{setup}, the model equations (\ref{model}-\ref{rmodel}) may be seen as containing important simplifications. Indeed, the modulator does not act as a frequency shifter, but rather creates sidebands around the two optical frequencies, so that at least six different optical frequencies are injected into each laser at the end of the feedback loop. However, only two of them are resonant: the self-feedback field, and one sideband of the other laser's field. Thus, we make the hypothesis that these resonant contributions dominate reinjection dynamics, and that at first order the nonresonant contributions can be neglected. In this respect, all the numerical simulations and the experiments were done for very small detuning values $\delta$, in the range $10^{-3} - 10^{-2}$. This means that in all the cases under study the fields that we have included in the model are very close to resonance, while the neglected terms are way out of resonance. We have checked our assumption by integrating a full model including the six feedback terms in a few typical cases, and have found that the non-resonant terms have no noticeable effect. Yet, this check cannot be done for all the cases presented in the following, because the time required for numerical integration becomes roughly 10 times longer for the full model. However, the overall good agreement between the experimental results and our simplified model makes us confident that the essential features are captured by the resonant terms only, and that our model is able to explain the characteristics of the systems, without resorting to more complicated and time-consuming simulations. 

\par Despite the complex reinjection scheme, we have already demonstrated experimentally that this setup can be used to lock the frequency difference of the lasers on the external synthetizer, \hl{resulting in a reduction of the beat-note linewidth from 600 kHz to less than 1 Hz}, with a locking range as large as $1~\mathrm{GHz}$~\cite{wang2014}. Nevertheless, the locking dynamics and the influence of the numerous experimental parameters had not been studied in detail, which motivates the comparison between the experimental observations and the rate equation model.

\subsection{Parameters of the model}

\par Following the setup described in Fig.~\ref{setup}, we focus on a particular form of the $\kappa_{ij}$ coefficients. First, this form has to take into account the asymmetric losses $Q_j$ of the on-chip coupler between the two lasers, which cause their output power to differ by $q = Q_2/Q_1 = I_2/I_1 = 0.25$ when they are pumped identically. Second, it must include the modulation ratio $m$ and the amplifier gain $G$. This leads to :
\begin{equation}
\begin{split}
\kappa = \begin{pmatrix}
\kappa_{11} & \kappa_{12} \\
\kappa_{21} & \kappa_{22}
\end{pmatrix} &= G \begin{pmatrix}
Q_1\sqrt{1-m} & \sqrt{Q_1 Q_2}\sqrt{m/2} \\
\sqrt{Q_1 Q_2}\sqrt{m/2} & Q_2\sqrt{1-m}
\end{pmatrix} \\ & =\kappa_0 \begin{pmatrix}
\sqrt{1-m} & \sqrt{qm/2} \\
\sqrt{qm/2} & q\sqrt{1-m}
\end{pmatrix}
\end{split}
\end{equation}

with $\kappa_0=GQ_1$ being the injection strength parameter. It can be controlled by the in-loop amplifier. However its absolute value depends on losses in the coupler and in the fiber injection, and cannot be measured precisely in our setup.

\par Among the system's parameters, we focus on the influence of detuning $\delta$, modulation ratio $m$, injection strength $\kappa_0$, and delay $\tau$. The other parameters  have been carefully measured for the DFB lasers and are kept fixed through the numerical study. Namely, by measuring relaxation oscillation frequencies and the lasers' transfer function under pump modulation, we conclude that $\tau_p=8\pm1~\mathrm{ps}$ and $\tau_c=60\pm18~\mathrm{ps}$, which leads to $\varepsilon' \approx 0.26$. Relaxation oscillations of the least pumped laser are $f_R^{(1)}\approx 8 ~\mathrm{GHz}$. Using a method based on optical injection~\cite{hui1990}, we also measured a relatively low Henry factor $\alpha=1.0\pm0.3$. In order to obtain a frequency difference $\nu_1-\nu_2$ around $10~\mathrm{GHz}$, we used $r_1=3$ and $r_2=4$. The fiber length of the whole feedback loop was $L\approx16$ m, \hl{which allows for coherent feedback and injection,} but gives a large normalized delay $\tau=4000$.

\begin{figure}[h!]
	\centering
	\vspace{-.2cm}	
	\includegraphics{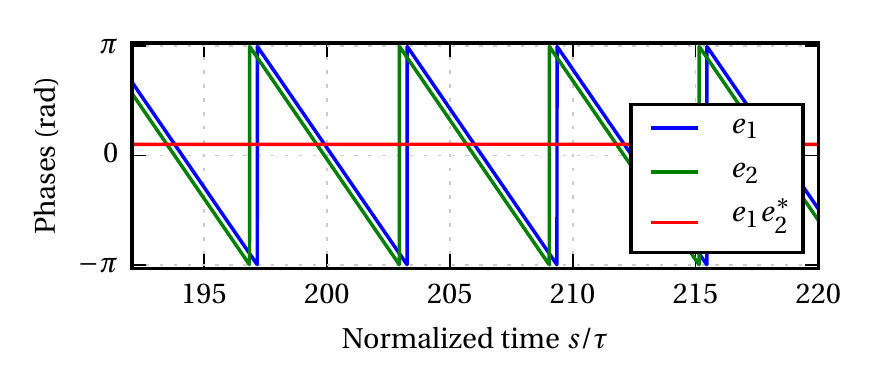}
	\vspace{-.8cm}	
	\caption{Time evolution of the phase of the two fields, obtained by numerical integration of Eqs.~(\ref{rmodel}). Their difference is constant, which corresponds to synchronized external cavity modes. Parameters are $\delta=0.002$, $\kappa_0=0.01$, $m=0.8$. and $\varphi_{1,2,x}=0,0,0$.}
	\label{ecm}
\end{figure}

\par Numerical integrations have been performed using the RADAR5 code~\cite{guglielmi2005}, and show that stable phase locking can indeed be achieved with this system. Figure~\ref{ecm} shows a typical case in which the two lasers settle on a stable external cavity mode (ECM), such that $e_1=|e_1|e^{i\omega s}$ and $e_2=|e_2| e^{i\gamma} e^{i\omega s}$. As with classical Lang-Kobayashi equations, this is expected to happen as equations~(\ref{rmodel}a) and (\ref{rmodel}b) feature a rotational symmetry $e_j\rightarrow e_je^{i\phi }$, with the same phase $\phi$ for the two fields~\cite{green2009}.

In the following study, we will be interested in the measurable outputs of the system, i.e. the characteristics of the beatnote. One is its amplitude defined as  $X=2\left|e_1e_2^*\right|$, and the other its phase with respect to the reference. Up to an additive constant, it corresponds to the phase difference between the two lasers $\theta=\arg{\left(e_1e_2^*\right)}$. When the frequency difference of the two lasers is stabilized, i.e. locked on the reference, this translates to a phase locking of the two lasers, so that their phase difference $\theta$ becomes constant.

\section{Numerical and experimental results}
\subsection{Feedback phases dependency}

\begin{figure}[h!]
	\vspace{-.2cm}
	\includegraphics{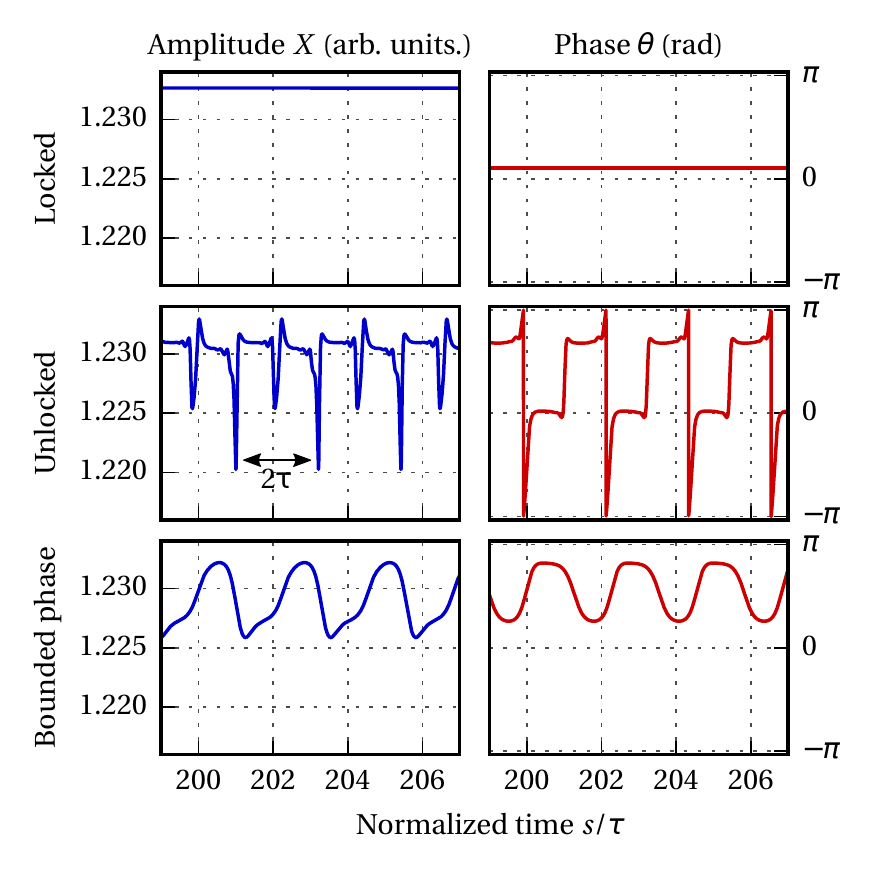}
	\vspace{-.8cm}	
	\caption{Example of very different long-term phase and beatnote intensity behaviors obtained by choosing different feedback phases. In the three cases $\delta=0.002$, $\kappa_0=0.01$, $m=0.8$. Locked : $\varphi_{1,2,x}=0,0,0$. Unlocked $\varphi_{1,2,x}=0,-1.3,2$. Bounded phase $\varphi_{1,2,x}=-2,-2.16,1$.}
	\label{phasedepexample}	
\end{figure}

Equations~(\ref{rmodel}a)--(\ref{rmodel}b) show that the three uncontrolled parameters $\varphi_j$ can play a role in the dynamics. Indeed, as seen in Fig.~\ref{phasedepexample}, slightly different feedback phases can lead to very different stable dynamics, from phase locking to $2\tau$-periodic oscillations, often found in delayed systems~\cite{dong2017}. Also bounded phase behavior can be found, as it is often the case in injected systems~\cite{kelleher2012}. In this case, phase oscillations are small enough so that their mean value remains constant, and the mean frequency difference is effectively locked on the reference oscillator, while the beatnote signal features amplitude and phase oscillations~\cite{romanelli2014}. 

The dependency of the dynamics on $\varphi_{1,2,x}$ may be detrimental if one is interested in obtaining stable, low noise opto-RF oscillators for microwave photonics applications, because optical phases can be difficult to control in an all-fibered setup, and are subject to environmental drifts. In order to evaluate the sensitivity to the feedback phases, we have performed numerical integrations with 50 different feedback phase values, taken from Halton sequences on $[-\pi,\pi]^3$. \hl{In order to take into account multistability, that is present in our system and is typical of nonlinear coupled systems involving large delays}, 5 different initial conditions were used for each phase triplet. For each integration, the phase difference in the steady state (for $s>10^7$) was computed from $e_1e_2^*$, and its extrema were recorded. The mean value of $\max\theta-\min\theta$ for all these points is shown on the maps in Fig.~\ref{phasedepmap}.

\begin{figure}[h!]
	\centering
	\vspace{-.2cm}	
	\includegraphics{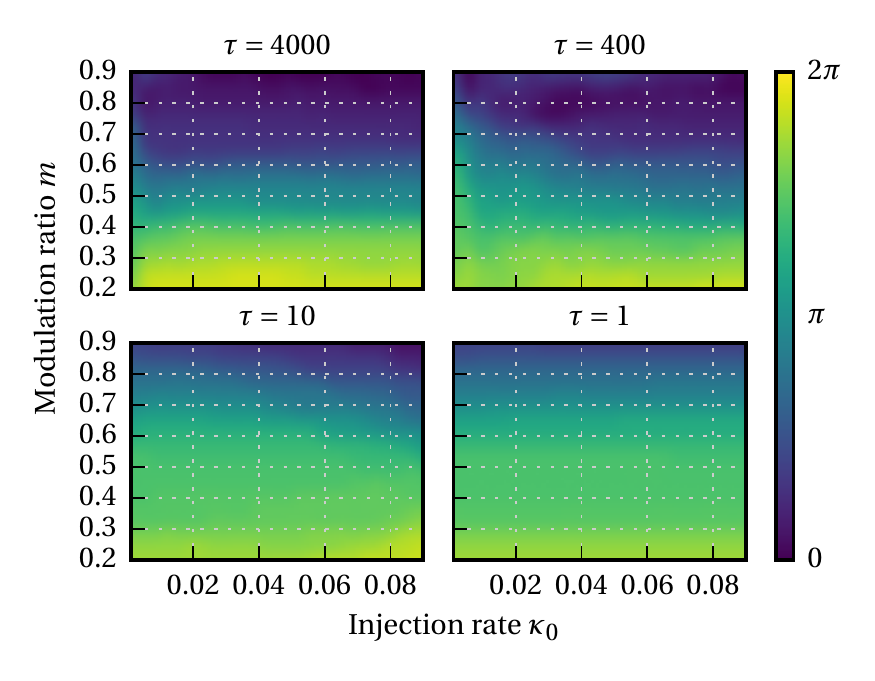}
	\vspace{-.8cm}	
	\caption{Mean value of the difference on phase extrema $\max\theta-\min\theta$, averaged on 50 different feedback phases values $\varphi_{1,2,x}$. Here we have taken $\delta=0$.}
	\label{phasedepmap}	
\end{figure}

We notice that the mean value never reaches zero, which means that it always exists a non-empty subset of the $(\varphi_1,\varphi_2,\varphi_x)$-space that leads to an unlocked regime. However, the volume of this subset becomes very small, down to about $3\%$ of the feedback phase space when using a high modulation ratio $m\approx0.8$. For $\tau=400$ a higher injection level $\kappa_0$ is needed to mitigate phase dependency. When the delay becomes smaller, for $\tau=10$, a very high modulation rate $m\approx0.9$ is needed to reach the same levels. This suggests the system becomes very sensitive to self-feedback, and that this should be avoided; a large delay seems to have a stabilizing effect. 
We also notice that the influence of $\kappa_0$ seems to be quite weak in the $10^{-2}$--$10^{-1}$ range.

\subsection{Absolute locking range and bounded phase dynamics}

In the case of optical injection between lasers with frequency detuning, one is interested in the locking range, i.e. how large can grow the detuning before the stable locking is broken. For low values of the Henry factor $\alpha$, the locking range usually grows with injection strength. Here we have also to take into account the delay and the feedback phases. 

\begin{figure}[h!]
	\centering
	\vspace{-.2cm}	
	\includegraphics{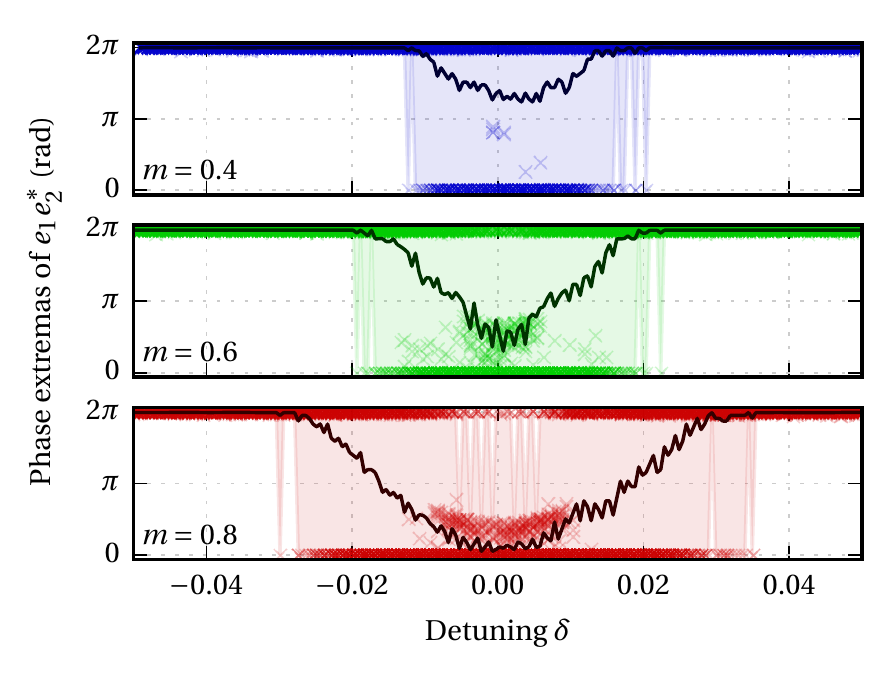}
	\vspace{-.8cm}	
	\caption{Numerical integrations for $\kappa_0=0.04$, $\tau=4000$ with varying detuning $\delta$. Each point corresponds to different feedback phases $\varphi_j$ and initial conditions, and the reported value is the stationary difference of output phase extrema $\max{\theta}-\min{\theta}$. Solid line shows the corresponding mean value for each $\delta$, and filled region helps seeing the range where there is at least one phase combination that makes locking possible.}
	\label{lockingranges}	
\end{figure}

Similarly to previous subsection 3.1, numerical simulations have been performed with 50 different values of $\varphi_j$ and varying $\delta$, and the results can be seen in Fig.~\ref{lockingranges}. Again, we notice that the dependency on the feedback phases is greatly reduced for higher $m$, but also that the region for which locking is possible is enhanced. We stress that this happens at constant $\kappa_0$, which means that the total injected intensity is the same, but is more balanced toward cross-injection than self-feedback.

Also, it can be noted that a lot of points do not lie on the $0$ or $2\pi$ line. They correspond to an oscillating output phase $\theta$ that remains bounded in the $[0,2\pi]$ interval, that is bounded phase dynamics, already seen on Fig.~\ref{phasedepexample}. Here we see that this feature is more prominent with high modulation rates $m>0.5$.

\begin{figure}[h!]
	\centering
	\vspace{-.2cm}	
	\includegraphics{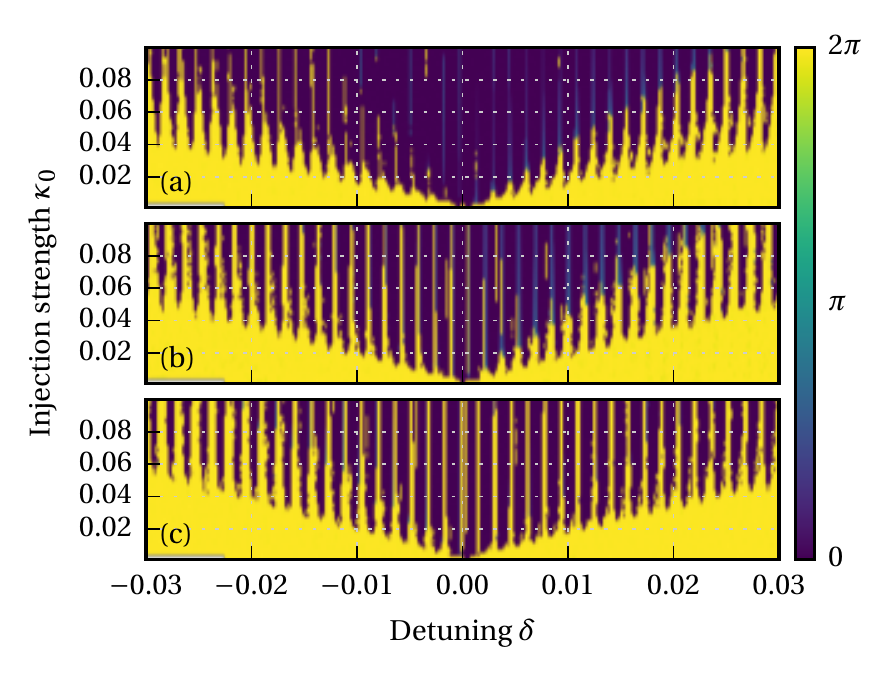}
	\vspace{-.8cm}
	\caption{Locking bands appear when detuning changes on a short time scale so that feedback phases can be considered constant. Other parameters are $m=0.8$ and $\tau=4000$. The three panels show different behaviors for different feedback phases : $\varphi_{1,2,x}=(-1.77, -0.97,  0.68)$ for (a), $( 0.88, -0.74 , -0.075)$ for (b) and $(-0.69,  1.36,  1.18)$ for (c).}
	\label{lockingbands}	
\end{figure}

\subsection{Locking bands}

Previous computations with random $\varphi_j$ (Fig.~\ref{lockingranges}) intended to stress the fact that optical feedback phases, that may not be controlled precisely, can alter the stability of the locking. Conversely, we can assume fixed $\varphi_j$, and see how the locking changes with the detuning $\delta$. Indeed, we see from equations (\ref{rmodel}b) and (\ref{phases}c-d) that the two injected terms include an additional detuning-dependent phase $e^{-i\delta\tau}$. As seen on Fig.~\ref{lockingbands}, this effective feedback phase variation breaks the locking range into periodic locking bands, with periodicity $2\pi/\tau$. However, the shape of the bands, and how they vary with $\kappa_0$ still depends on the optical phases $\varphi_j$. Between the stable bands, and depending on the other phases, we observe either complete unlocking or bounded phase with $2\tau$-periodic output.

\subsection{Experimental results}

\begin{figure}[h!]
	\centering
	\includegraphics{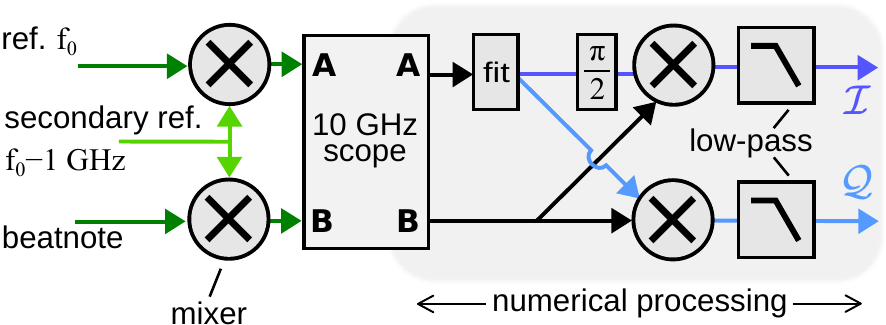}
	\caption{\hl{Experimental apparatus used for the demodulation. The beatnote signal and the reference signal are down-converted at 1~GHz. The time series are then numerically processed. First, the downconverted reference is fitted with a sinusoidal waveform to remove noise. Then, numerical multiplication and filtering gives the signal quadratures $\mathcal{I}$ and $\mathcal{Q}$, which are used to retrieve the amplitude and phase.}}
	\label{demod}	
\end{figure}

\begin{figure}[h!]
	\centering
	\vspace{-.2cm}	
	\includegraphics{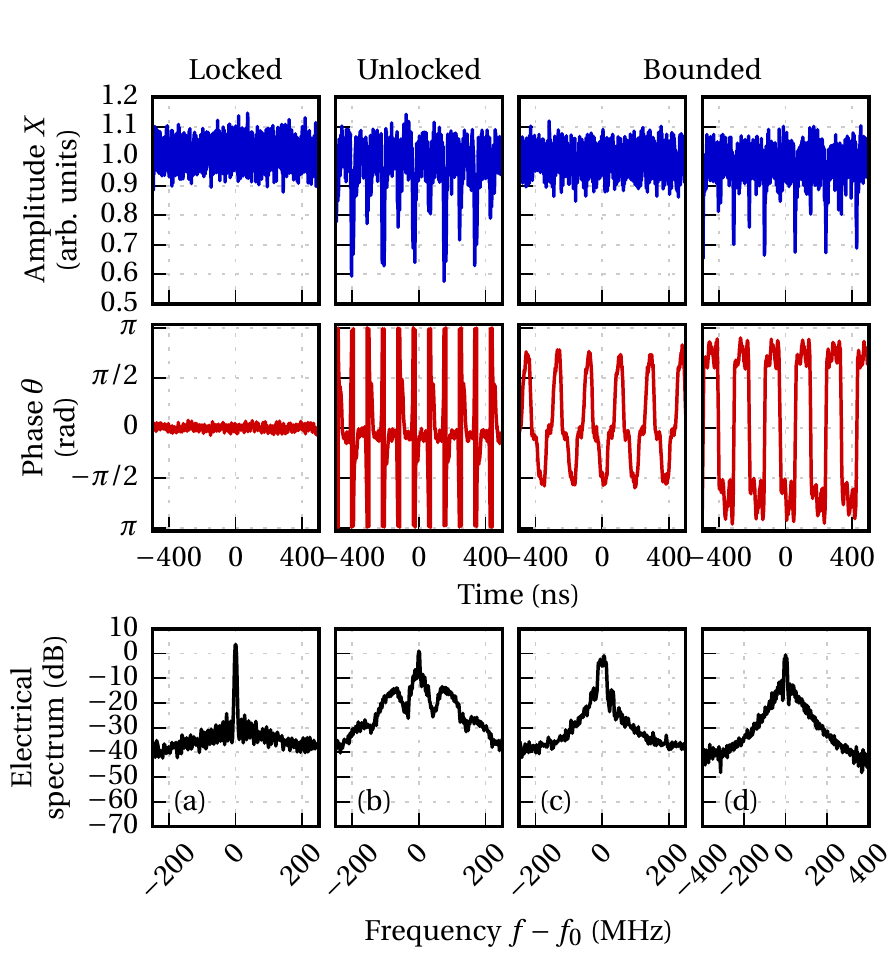}
	\vspace{-.8cm}	
	\caption{Experimental time series and electrical spectra. The free-running beat-note is at $10~\mathrm{GHz}$, and the relaxation oscillations are $f_R^{(1)}=8~\mathrm{GHz}$. We set $m=0.8$, and the detuning $\nu_1-\nu_2-f_0$ are respectively $123~\mathrm{MHz}$, $29~\mathrm{MHz}$, $-26~\mathrm{MHz}$ and $174~\mathrm{MHz}$.}
	\label{exptraces}	
\end{figure}

The regimes previously discussed can be observed experimentally. As an illustration, some time series of the beat-note signal observed on a photodiode are presented on Fig.~\ref{exptraces}. In order to obtain them, the beat-note signal and the reference at $f_0$ were downconverted at around $1~\mathrm{GHz}$ by mixing them with a local oscillator, and recorded by a fast oscilloscope. Then, the phase and amplitude time series have been obtained by demodulating the downconverted beatnote with the downconverted reference. See Fig.~\ref{demod} for the full setup. As compared with numerical results, a range of behaviors from locked state (a), 
to $2\tau$-periodic dynamics, either with bounded (c)--(d), or unbounded phase (b) are found, in qualitative agreement with the simulations. The comparison with numerics is easier to be done for the phase time series; on the contrary, since the beatnote signal was very weak on the oscilloscope, the resulting amplitude signal is quite affected by the noise floor due to the detection electronics. When the frequency difference is locked on the reference, the spectral purity of the synthetizer is completely transferred on the beatnote spectrum (Fig. \ref{exptraces}a). This can also be seen on phase noise measurements, as reported in~\cite{wang2014}, {where the phase noise of output beatnote matches the one of the reference for offset frequencies lower than the external cavity free spectral range.} \hl{In the bounded phase regime, a larger electrical spectrum is observed (Fig.~}\ref{exptraces}\hl{c--d), as it includes a number of harmonics related to the small oscillations of the phase and amplitude.}

\begin{figure}[h!]
	\centering
	\vspace{-.2cm}	
	\includegraphics{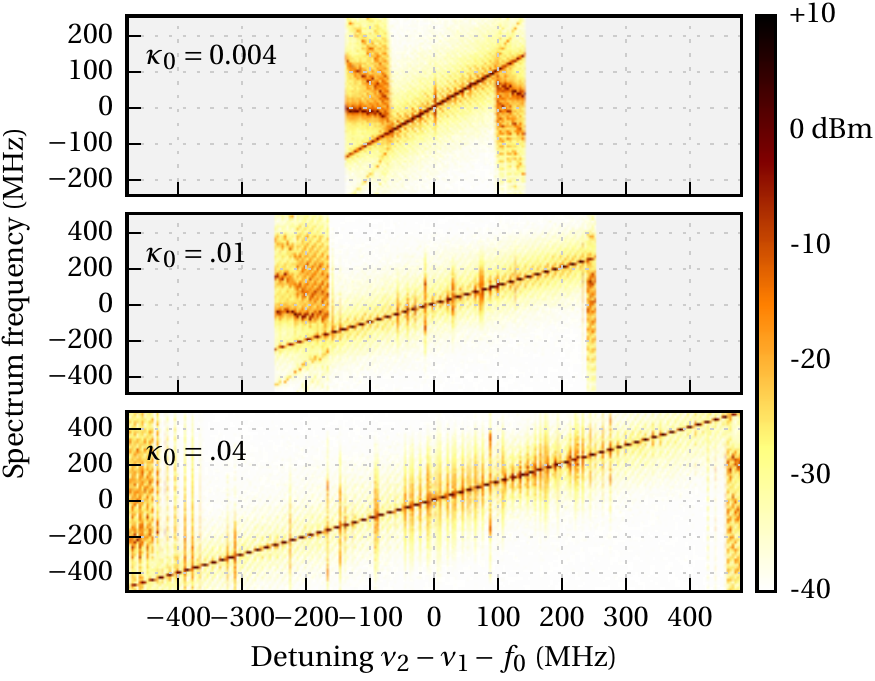}
	\vspace{-.8cm}	
	\caption{Experimental spectrograms as a function of the detuning. Modulation ratio is set to maximum $m\approx0.9$. For different injection strength $\kappa_0$, the control frequency $f_0$ is swept around the free-running beatnote $\nu_2-\nu_1$. The electrical spectrum is recorded at each point, and plotted vertically.}
	\label{spectrograms}	
\end{figure}

We then varied the detuning $\delta$ in small steps, and recorded the beatnote electrical spectrum as well as time series. On Fig.~\ref{spectrograms}, the locking bands expected from the numerical simulations appear clearly. This fact suggests that the feedback phases $\varphi_j$ vary indeed on slower time scales than the measurement time, i.e. a few minutes. If they were to vary faster, the locking bands would not be observed, or be very blurred, as we know from Fig.~\ref{lockingbands} that their position depends on the feedback phases. Locking regions, where the electrical spectrum features only one peak at $f_0$, alternate with unlocking zones, where a more complicated spectrum can be seen. The spacing of $1/T=12.5~\mathrm{MHz}$ between the locking bands is also observed\hl{, as well as the asymmetry with respect to the detuning. This asymmetry, due to $\alpha$, appears also in the simulations of Fig.~7}. As in Fig.~\ref{lockingranges}, the locking range depends on the injection strength $\kappa_0$, and can reach roughly $800~\mathrm{MHz}$. The experimental value of $\kappa_0$ is roughly estimated using a simple injection experiment, but it gives results in agreement with simulations. For instance, the observed locking range of $800~\mathrm{MHz}$ for $\kappa_0\approx0.04$ corresponds to $|\delta|<0.05$, which is a locking range in the correct order of magnitude according to Fig.~\ref{lockingranges}.

\begin{figure}[h!]
	\centering
	\vspace{-.2cm}	
	\includegraphics{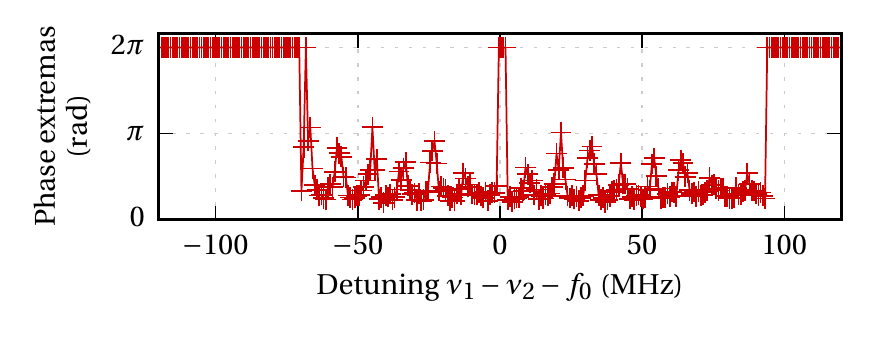}
	\vspace{-.8cm}	
	\caption{Phase extremas with varying detuning, for $\kappa_0=0.04$ and $m=0.8$. The phase does not go to zero because of experimental noise.}
	\label{expphase}	
\end{figure}

Although electrical spectra allow to discriminate between locked and non-locked regions, they do not give any information about the phase dynamics between the locking range. A more detailed view of the beatnote phase extrema can be seen on Fig.~\ref{expphase} for $\kappa_0=0.04$. We clearly see locking bands where $\max\theta-\min\theta=0$, bounded phase regions, and an unbounded zone at~$\delta\approx0$. 

\begin{figure}[h!]
	\centering
	\vspace{-.2cm}	
	\includegraphics{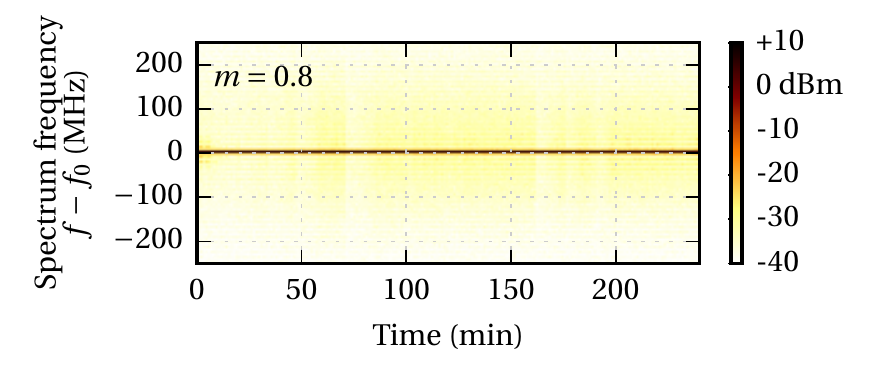}
	\vspace{-.8cm}	
	\caption{Spectrogram of the beat-note frequency, showing the long-term stability of the locking.}
	\label{stability}	
\end{figure}

Finally, we turned to the issue of the long-term stability of the phase-locking of the optical beat-note on the synthesizer. 
Fig~\ref{stability} displays a spectrogram of the beat-note frequency for $m\approx0.8$. In that case, frequency locking was observed for at least 12 hours. We can give a rough estimation of the variation of the feedback phases during this time interval. Indeed, using a high-resolution (5 MHz) optical spectrum analyzer, we have measured the optical frequencies of the free-running lasers over two hours. 

\begin{figure}[h!]
	\centering
	\vspace{-.2cm}	
	\includegraphics{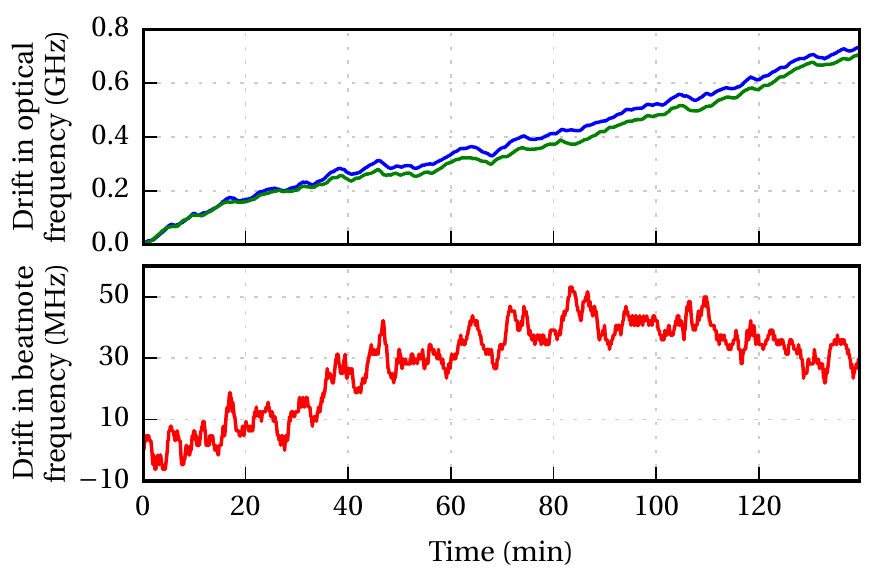}
	\vspace{-.8cm}	
	\caption{Long-term drift of the optical frequencies of the free-running lasers (upper panel), and of their difference (lower panel).}
	\label{fig:opticalfreq}	
\end{figure}

As can be seen from Fig.~\ref{fig:opticalfreq}, the optical frequencies exhibit strongly correlated drifts, thanks to the monolithic structure, of about 600 MHz in two hours, i.e. 80 kHz/s. Recalling the expressions (\ref{phases}) of the optical feedback phases, and taking ${T \sim 100~\mathrm{ns}}$, we get for instance a variation of the self-feedback phase $\varphi_1$ for the first mode $\delta \varphi_1 \sim 2 \pi$  8 mrad/s. This means that a $2 \pi$ variation of $\varphi_1$ takes $\sim 2$ min. $\varphi_2$ and $\varphi_x$ experience similar drifts, which are, however, strongly correlated to $\delta \varphi_1$.
Indeed, $\varphi_2$ can be written as

\begin{equation}
\varphi_2 = \varphi_1 + 2\pi \frac{(\nu_1-\nu_2) nL}{c} + \psi_{22}-\psi_{21}
\end{equation}  

Thus, if we ignore the phases $\psi_{ij}$, the variation of $\varphi_2$ with respect to $\varphi_1$ is ruled by the drift of the frequency difference $\nu_1 - \nu_2$ , which, according to Fig.~\ref{fig:opticalfreq}, is ten times slower than the optical frequency drift, corresponding roughly to 60 MHz in two hours. This means that $\delta \varphi_2 - \delta \varphi_1$ takes $\sim$ 20 min to make a 2$\pi$ excursion.
It is difficult to establish a precise link between these drifts and the feedback phase sensitivity studied in Fig.~\ref{phasedepmap}: indeed, the above estimates only take into account the drift of the optical frequencies, but ignore environmental fluctuations included in $\psi_{ij}$; furthermore, as explained above, the feedback phases do not sample homogeneously the $[-\pi,\pi]^3$ cube, as their variations are strongly correlated. 
However, Figs.~\ref{stability} and \ref{fig:opticalfreq} demonstrate that stable phase locking can be maintained over several hours, in spite of uncontrolled feedback phase fluctuations that occur on the time scale of $\sim$ 10 min. 


\section{Discussion}

We have shown that two DFB lasers coupled by frequency-shifted cross-injection and self-feedback can feature stable frequency locking on a broad range of parameters. Even if their variations cannot be controlled, the optical feedback phases do not prevent the obtention of stable phase-locking, which is more robust against perturbations when using a strong modulation rate, i.e. favoring cross-injection over self-feedback. The large delay $\tau\approx4000$ associated to the feedback loop turns out to be preferable to a shorter one $\tau\approx1-10$. We also noted that such a large delay, along with a frequency detuning, breaks the locking range into locking bands, a phenomenon already observed in injection-feedback systems~\cite{nizette2004}. In-between the locking regions, complex outputs such as bounded phase synchronization and $2\tau$-periodic output have been observed experimentally and numerically.

We see that numerical simulations are in good agreement with the experimental observations. This suggests that the observed behaviors are quite robust and general in this kind of systems. Yet, very different values for some parameters could alter the global behavior, for instance a sensibly higher $\alpha$ factor would lead to a quite different picture. The dependence on the laser parameters has still to be precisely explored, as well as other coupling schemes, for instance with pure frequency shifting elements. {In that last case, it has been shown that stabilization is also possible}~\cite{wang2014}. 
\hl{The technique presented here allows continuous frequency tuning and can straightforwardly be extended to higher frequencies without degradation of the phase noise. In principle, the only limitation is set by the bandwidth of the modulator. For instance, 90~GHz signal generation has been demonstrated by using a higher frequency difference between the lasers and fast modulators}~\cite{kervella2015}.
It could also be used on dual-polarization semiconductor lasers, for instance VECSELs~\cite{baili2009}, where polarization allows for a finer control of injection.

Finally, the improved theoretical understanding provided by the present work will assist the design of novel opto-electronic oscillator schemes. 
\hl{Indeed, in an alternative architecture it is possible to remove the electronic synthesizer, and use the beatnote signal received by the photodiode to drive the MZM, thus obtaining a self-referenced hybrid opto-electronic oscillator with single-sideband output}~\cite{vallet2016,thorette2018}.   
{Also, it has started being used on integrated single components}~\cite{primiani2016}. We expect that the present study will provide insights towards better stability and phase noise performances for microwave photonics signal generation.
We thank F. van Dijk from III-V Labs for providing  the DFB lasers, M. Guionie for measurements on the lasers, S. Bouhier for technical assistance on electronics and G. Raffy for support on the computation cluster. This work was partially funded by the European Defense Agency (EDA) through the HIPPOMOS project.

()


\end{document}